# Robust End-to-End FSO Transmission with Joint Coding Modulation and BiLSTM-Based Channel Modeling under Atmospheric Turbulence

Wei Zhang, Zhenming Yu, *Member, IEEE*, Hongyu Huang, Xiangyong Dong, Yan Ma, Yongli Zhao, Shanguo Huang, and Kun Xu

*Abstract*—Free space optical (FSO) communication is considered a promising solution in next-generation communication networks. However, its performance is significantly influenced by atmospheric turbulence. To enhance system robustness to turbulence, we propose a turbulence-robust end-to-end FSO communication system (TRFSO) that integrates a data-driven channel model with joint source–channel coding modulation (JSCCM). Specifically, a bidirectional long short-term memory (BiLSTM)-based channel model is developed and trained on data collected over a physical FSO link under varying turbulence conditions. This model accurately captures real-world channel distortions, achieving a minimum Kullback–Leibler (KL) divergence of 0.0019 in amplitude distribution matching. Experimental results show that the TRFSO system trained with the BiLSTM-based channel model outperforms the same architecture trained under the additive white Gaussian noise (AWGN) channel, achieving an average 3.5 dB improvement in multi-scale structural similarity (MS-SSIM) under strong atmospheric turbulence. These results demonstrate the effectiveness of the proposed TRFSO in achieving robust and reliable transmission under dynamic atmospheric turbulence.

*Index Terms*—Free space optical communication, BiLSTM, joint source–channel coding modulation.

## I. INTRODUCTION

FREE space optical (FSO) communication has attracted increasing attention as a key technology for next-generation networks, owing to its advantages of high bandwidth, low latency, strong security, and strong resistance to electromagnetic interference [1-3]. With the growing demands of future 6G, FSO is considered a promising solution for high-capacity data transmission in satellite-to-ground, inter-satellite, and terrestrial communication scenarios [4-5].

Despite its great potential, the deployment of FSO systems remains highly susceptible to atmospheric turbulence, which results from random fluctuations in the refractive index of air due to temperature and pressure variations [6]. Such turbulence introduces amplitude fading, phase distortions, and beam wandering, severely degrading the quality and reliability of optical links [7]. To mitigate these impairments, several physical-layer techniques have been investigated. Adaptive optics (AO) has been widely adopted as a solution to correct wavefront distortions in real time using deformable mirrors and wavefront sensors, thereby enhancing beam quality and reducing signal fading [8]. However, AO systems often involve high hardware complexity and high implementation cost, which limit their adaptability in resource-constrained environments [9,10].

In addition to hardware-based compensation approaches such as AO, turbulence-robust strategies based on joint source–channel coding modulation (JSCCM) have also been explored. For example, the discrete-time analog transmission scheme demonstrates improved adaptability to varying channel states compared with traditional digital transmission [11]. In contrast to conventional communication systems that follow a separate source and channel coding paradigm [12], JSCCM directly map source information into transmitted symbols. Although Shannon's separation theorem shows that the separate coding paradigm can achieve theoretical optimality, this result relies on the assumption of infinitely long source and channel blocks, which is difficult to satisfy under practical constraints such as limited bandwidth and low latency [13-15]. Recently, the paradigm of JSCCM has shown promising results in both wireless and optical communication scenarios [16-20]. Nevertheless, most existing JSCCM models are trained on additive white Gaussian noise (AWGN) channel, which fail to capture the non-Gaussian characteristics of real-world FSO channel. Consequently, their performance degrades in practical deployments.

To bridge the gap between theoretical design and real-world deployment, accurate and differentiable channel modeling is essential for enabling robust end-to-end training in FSO communication systems. Existing models can generally be

Manuscript received × × ××; revised × × ××. This work was financially supported by the National Key R&D Program of China (No.2023YFB2905900); National Natural Science Foundation of China (No. 62522502, 62371056); Major Science and Technology Support Program of Hebei Province (No. 252X1701D); Sponsored by Beijing Nova Progra; Shenzhen Science and Technology Program (KJZD20230923115202006); the Fund of State Key Laboratory of Information Photonics and Optical Communication BUPT (No. IPOC2025ZZ02); the Fundamental Research Funds for the Central Universities(No. 530424001, 2024ZCJH13). (Corresponding author: Zhenming Yu.)

Wei Zhang, Zhenming Yu, Hongyu Huang, Xiangyong Dong, Yan Ma, Yongli Zhao, Shanguo Huang, and Kun Xu are with State Key Laboratory of Information Photonics and Optical Communications, Beijing University of Posts and Telecommunications, Beijing, China. (e-mails: zw_bupt@bupt.edu.cn; yuzhenming@bupt.edu.cn; hongyuhuang@bupt.edu.cn; dongxiangyong@bupt.edu.cn; 2024010208@bupt.edu.cn; yonglizhao@bupt.edu.cn; shghuang@bupt.edu.cn; xukun@bupt.edu.cn).

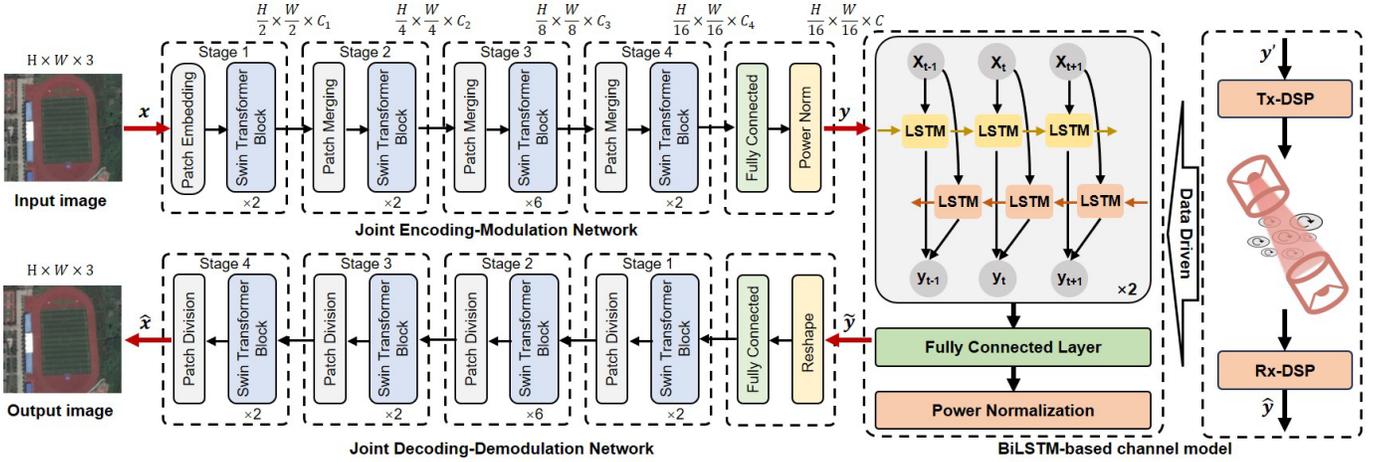

**Fig. 1.** Proposed structure of the turbulence-robust end-to-end FSO communication system. Tx-DSP: digital signal processing at the transmitter; Rx-DSP: digital signal processing at the receiver.

divided into two categories. The first comprises statistical channel models, such as log-normal and gamma-gamma fading distributions [1]. Although these models can approximate the statistical behavior of turbulence-induced fading, they are typically based on simplifying assumptions and do not account for system-level impairments. In addition, their non-differentiable nature makes them incompatible with gradient-based optimization in deep learning frameworks. The second category includes data-driven models based on deep learning [21,22]. These models are differentiable but are trained on simulation-generated datasets and focus on the optical channel layer, without incorporating source encoding or decoding processes. Consequently, they are not suitable for end-to-end optimization and tend to exhibit degraded performance when deployed in real-world optical systems.

To address these challenges, we propose a turbulence-robust end-to-end FSO communication system (TRFSO). The TRFSO integrates a data-driven channel model based on bidirectional long short-term memory (BiLSTM) networks with JSCCM architecture built on Swin Transformer blocks. The proposed BiLSTM-based channel model is trained on signal pairs collected from a physical FSO transmission link under varying turbulence conditions. Its output distribution closely matches that of the real channel, achieving a minimum Kullback–Leibler (KL) divergence of 0.0019, demonstrating high modeling fidelity. We build a 30 Gbaud FSO experimental platform employing intensity modulation/direct detection (IM/DD), where a spatial light modulator (SLM) is used to experimentally emulate atmospheric turbulence with adjustable intensity. Based on this setup, we systematically compare the performance of the proposed TRFSO with baseline transmission schemes under various turbulence strengths, ground station altitudes, and zenith angles. Experimental results show that the proposed TRFSO improves communication stability and image transmission quality. Under strong turbulence, it achieves an average 3.5 dB improvement in multi-scale structural similarity (MS-SSIM) compared to TRFSO trained under the AWGN channel (TRFSO-AWGN), without the BiLSTM-based channel model.

This paper is organized as follows. Section II describes the design and architecture of the proposed TRFSO. Section III presents the experimental setup. Section IV presents and analyzes the experimental results. Finally, in the concluding section, we summarize our approach and results.

## II. SYSTEM ARCHITECTURE

In this section, we introduce the architecture of the proposed TRFSO system, including the JSCCM framework and the BiLSTM-based channel model. We also outline the adopted training optimization strategy.

### A. Overview of the TRFSO

As shown in Fig. 1, the proposed TRFSO system consists of three main components: a joint encoding–modulation (JEM) network, a BiLSTM-based channel model, and a joint decoding–demodulation (JDD) network. The entire system is trained end-to-end, enabling the JEM and JDD to adapt to FSO channel impairments. At the transmitter side, the JEM maps the input image $x$ into symbols $y$. The JEM network is constructed using a hierarchical Swin Transformer backbone, which captures both local and global semantic features at multiple resolutions. During the training phase of TRFSO, the encoded symbols $y$ are passed through the BiLSTM-based channel model, which acts as a differentiable replacement for the real FSO transmission link, enabling gradient-based end-to-end optimization. In the testing phase, the encoded symbols are directly transmitted through the actual FSO channel, bypassing the learned channel model. On the receiver side, the JDD reconstructs the output image $\hat{x}$ from the received noisy symbols. The JDD mirrors the structure of the JEM, progressively recovering spatial resolution and image content. The overall architecture supports differentiable training, allowing gradient backpropagation through the entire system, including the channel model. This makes it possible to optimize the JEM and JDD jointly under FSO distortions.

## B. Design of the JEM and JDD

To enable efficient compression and reconstruction, the proposed TRFSO adopts Swin Transformer as the core backbone for both the JEM and JDD. Its hierarchical structure with shifted window self-attention captures long-range dependencies and multiscale features with linear computational complexity [23]. Compared to convolutional neural networks (CNNs), Swin Transformer offers stronger global context modeling, which is especially beneficial for image transmission. This capability allows the encoder–decoder to extract compact semantic representations and reconstruct high-quality images under channel distortions.

The JEM first partitions the input image $x \in R^{H \times W \times 3}$ into non-overlapping patches with a patch size of $2 \times 2$. Each patch is them mapped to a feature vector of dimension $C_1 = 128$ through a linear embedding layer. The embedded tokens are subsequently processed by two Swin Transformer blocks, collectively forming the "stage 1" of the JEM. The next three stages each begin with a patch merging layer that reduces the spatial resolution by a factor of 2 while increasing the dimension from $C_1$ to $[C_2, C_3, C_4] = [192, 256, 320]$, respectively. These four stages contain 2, 2, 6, and 2 Swin Transformer blocks, respectively. After the final stage, the output is passed through a fully connected (FC) layer that maps the high-dimensional features to the desired number of transmission dimension $C$, determined by the compression ratio. In our work, $C$ is set to 96. A power normalization layer is then applied to ensure a fixed average transmission power across all symbols.

The JDD mirrors the JEM architecture in reverse. It receives the noisy symbol and reshapes it to $\frac{H}{16} \times \frac{W}{16} \times C$. An FC layer then maps it back to a high-dimensional latent space of size $\frac{H}{16} \times \frac{W}{16} \times C_4$. The JDD consists of four stages, each employing patch division layers to progressively upsample the spatial resolution while reducing channel dimensions from $C_4$ to $3$. These stages contain 2, 2, 6, and 2 Swin Transformer blocks, respectively, corresponding to the structure of JEM.

Given that high peak-to-average power ratio (PAPR) in transmitted signals can cause nonlinear distortion in physical FSO systems such as AD/DA and modulators, it is crucial to constrain the PAPR during training [24]. To address this, and in line with strategies explored in prior work [11], we design a customized loss function that jointly considers image reconstruction quality and a PAPR-aware term. Specifically, the system loss is defined as:

$$L_f = L_d(x, \hat{x}) + \lambda_{PAPR} \cdot PAPR(y), \qquad (1)$$

where $L_d(x, \hat{x})$ denotes the image distortion loss, and $PAPR(y)$ is computed from $y$ output by the encoder, and penalizes large power fluctuations to suppress nonlinear distortion in the physical transmission chain. In our work, the $\lambda_{PAPR}$ is set to $5 \times 10^{-4}$.

## C. BiLSTM-Based Channel Model

To accurately simulate the distortion characteristics of real-world FSO transmission, we design a data-driven channel model based on BiLSTM networks. BiLSTM is an extension of standard LSTM that processes sequence data in both forward and backward directions, allowing it to capture richer temporal dependencies. This capability is particularly well suited to model FSO systems. The proposed channel model is composed of two stacked BiLSTM layers, followed by an FC projection module and a power normalization layer.

Specifically, the output symbol sequence from the JEM is first reshaped into a 2D tensor of shape $N \times M$, where $M$ denotes the input size of the channel model and $N$ is the feature dimension. This reshaped sequence is then fed into two stacked BiLSTM layers with a hidden size of 30. To prevent overfitting and enhance generalization, a dropout layer with a dropout rate of 0.15 is inserted between the two BiLSTM layers. Due to the bidirectional nature of BiLSTM, the output has a feature dimension of 60 per time step (i.e., 30 units from the forward LSTM and 30 from the backward LSTM). The BiLSTM output is passed through an FC layer, which consists of two linear layers with neurons of 120 and 60, each followed by a LeakyReLU activation (with a negative slope of 0.01). A power normalization layer is applied at the end.

## D. Model Training

The training of the TRFSO system is carried out in three stages.

In the first stage, we pre-train the JEM and JDD networks under AWGN channel. This step provides a baseline model for data collection of the channel model and offers initial encoder–decoder weights. The JEM encodes input images $x'$ into symbols $y'$ for transmission. The symbols $y'$ are then passed through an FSO transmission system, consisting of the digital signal processing at the transmitter (Tx-DSP) module, an IM/DD-based FSO transmission link, and digital signal processing at the receiver (Rx-DSP) module. The received signal after Rx-DSP is denoted as $y''$.

In the second stage, the collected transmission results $y''$, together with the corresponding transmitted inputs $y'$, are used to form training pairs $(y', y'')$ for training the channel model. We use phase screens displayed on the SLM to simulate various levels of atmospheric turbulence. In total, six different noise levels are modeled to generate corresponding channel datasets. For each level, the first 120,000 symbols are used for training, and an additional 12,000 symbols are used for testing. The BiLSTM channel model takes $y'$ as input and outputs $\hat{y}$, an estimate of the received signal. The model is trained to minimize the mean squared error (MSE) between the predicted and actual outputs:

$$L_c = MSE(\hat{y}, y''). \qquad (2)$$

In the third stage, the parameters of the channel model are fixed, and it is embedded into the TRFSO system as a differentiable channel layer, replacing the AWGN model used in the first stage. The end-to-end training is then conducted, enabling the JEM and JDD networks to adapt to the distortion characteristics of real FSO channels. To improve the robustness of the model under varying atmospheric conditions, we adopt a training strategy that randomly samples from multiple channel models corresponding to different turbulence levels in each training iteration. This encourages the JEM and JDD to generalize across a wide range of FSO scenarios and improve the resilience of the system to dynamic atmospheric environments.

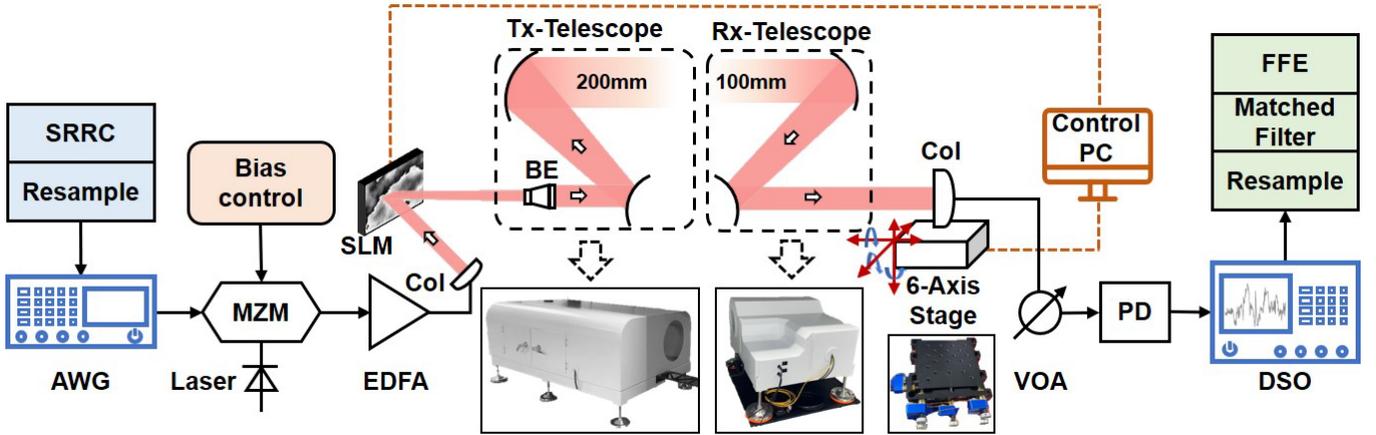

**Fig. 2.** Experiment setup. AWG: arbitrary waveform generator; MZM: Mach-Zehnder modulator; EDFA: erbium-doped fiber amplifier; Col: collimator; SLM: spatial light modulator; BE: beam expander; VOA: variable optical attenuator; PD: photodetector; DSO: digital storage oscilloscope.

## III. EXPERIMENTAL SETUP

To evaluate the performance of the proposed JSCCM-FSO and collect data for training the channel model, we build a FSO transmission system based on intensity modulation and direct detection (IM/DD), as illustrated in Fig. 2. The system operates at a symbol rate of 30 Gbaud.

At the transmitter, the symbols generated by JEM are shaped by a square-root raised cosine (SRRC) filter with a roll-off factor of 0.1 and then resampled to 60 GSa/s. The signals are then loaded into an arbitrary waveform generator (AWG, Keysight M8195A). This AWG offers a maximum sampling rate of 65 GSa/s and an analog bandwidth of 25 GHz. The electrical analog signals from the AWG are first amplified by a linear broadband amplifier (SHF S807C) and then fed into a Mach-Zehnder modulator (EOSPACE AX-0MVS-40-PFA-PFA-LV) to generate the optical waveform. The laser source (CoBrite-DX) launches an optical wave at 1550 nm with an optical power of 10 dBm and feeds it to the MZM. The optical signal is amplified by an erbium-doped fiber amplifier (EDFA, Amonics AEDFA-33-B-FA), then collimated into free space. To experimentally simulate atmospheric turbulence, phase screens were generated using the power spectrum inversion method and loaded onto the SLM [25]. For both strong and moderate turbulence conditions, 50 phase screens were created, respectively. The optical beam is expanded to a diameter of 200 mm by the transmitting telescope, matching the beam size expected from a divergence of 40 μrad over a 5 km free-space transmission.

At the receiver, the distorted optical beam is collected by a 100 mm telescope. The beam is then coupled into a standard single-mode fiber via an optical collimator mounted on a six-axis displacement stage. This stage uses six independent motors to control the position and orientation of the fiber collimator, enabling precise adjustments in translation, tilt, and rotation. Such fine-grained control is essential to ensure stable coupling efficiency. A variable optical attenuator (VOA) emulates transmission attenuation. The coupled optical signal is converted into the electrical domain using a high-speed photodetector (PD, FINISAR XPRV2325A) with a bandwidth of 30 GHz. The resulting electrical waveforms are sampled by a digital storage oscilloscope (DSO, Keysight UXR0594A). The sampled waveforms are then processed by the Rx-DSP, which includes resampling, matched filtering, and a 61-tap feedforward equalizer (FFE). The control PC controls both the SLM phase screen updates and the adjustment of the six-axis stage.

TABLE I
NUMBER OF SYMBOLS AFTER CODING AND MODULATION.

| Method | Number of symbols |
|---|---|
| JPEG2K+LDPC+OOK | 124416 |
| JPEG2K+LDPC+PAM4 | 122472 |
| TRFSO-AWGN | 122880 |
| TRFSO | 122880 |

We compare the proposed TRFSO system against two baselines: traditional FSO communication systems employing separate coding modulation and TRFSO-AWGN, which shares the same network architecture and training strategy as TRFSO but is trained under an AWGN channel instead of the proposed BiLSTM-based channel model. In the traditional FSO communication systems, JPEG2000 (JPEG2K) [26] is used for source coding, and a rate-3/4 LDPC code for channel coding [27]. The modulation formats applied are PAM4 and PAM8. With a fixed system baud rate, the compression rate of each transmission scheme directly affects the image transmission rate. For fairness, all schemes are configured to produce a comparable number of transmitted symbols after encoding and modulation. Specifically, for an image with resolution $512\times 640$, we calculate the total number of bits and symbols generated by each method, as summarized in Table I. We adopt the MS-SSIM as the metric to evaluate the perceptual quality of the reconstructed images after transmission [28]. MS-SSIM is widely used in image transmission tasks, as it better aligns with human visual perception compared to traditional metrics such as peak-signal-to-noise ratio (PSNR).

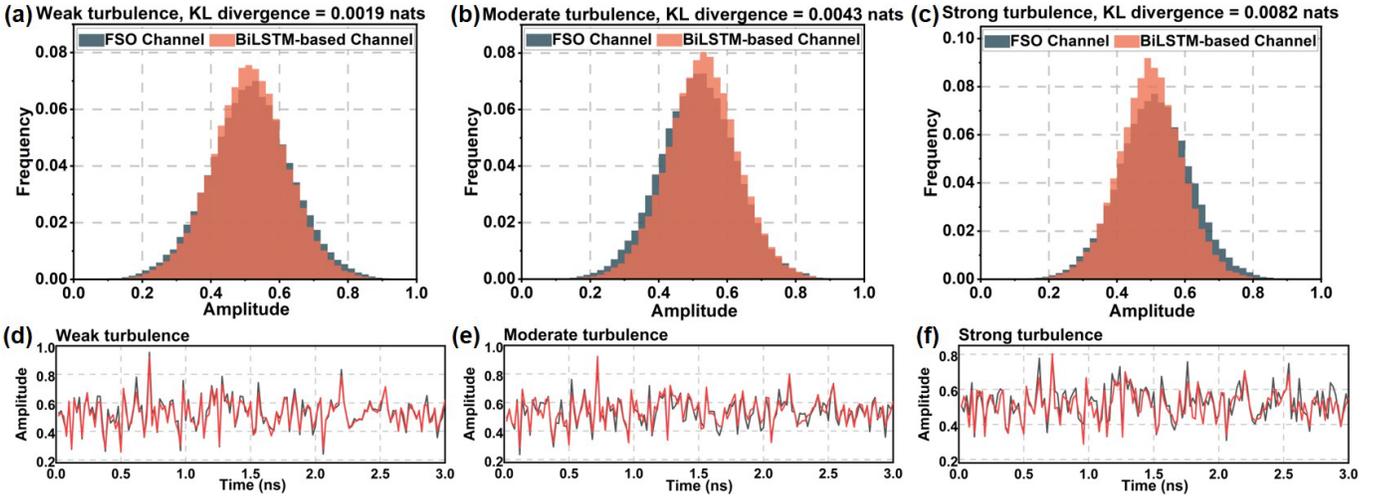

**Fig. 3.** Amplitude distribution and waveform comparison between the real FSO channel and the BiLSTM-based channel model under (a) (d) weak, (b) (e) moderate, and (c) (f) strong turbulence.

Both the proposed TRFSO and TRFSO-AWGN are trained on the DIV2K dataset, which provides 800 natural images of 2K resolutions [29]. For evaluation, we use the RSOD dataset, which contains remote sensing images with four object types: airplanes, oil tanks, playgrounds, and overpasses [30].

### IV. EXPERIMENTAL RESULTS

*A. Performance of the BiLSTM-based Channel Model*

To train the proposed BiLSTM-based channel model, we collected seven datasets under varying levels of atmospheric turbulence. The signal-to-noise ratios (SNRs) of these datasets, computed between the transmitted symbols $y'$ and the received symbols $y''$ after Rx-DSP, range from 3 dB to 13 dB. To validate the effectiveness of the proposed channel model, we compared its output distributions with those measured from the physical FSO link under three representative turbulence conditions, characterized by refractive index structure parameters $C_n^2$ of $10^{-17}$, $10^{-15}$, and $10^{-13}$ m$^{-2/3}$, respectively. The similarity between the modeled and measured amplitude distributions was quantified using the KL divergence. KL divergence measures the difference between two probability distributions, with lower values indicating higher similarity. As shown in Fig. 3(a)-(c), the BiLSTM-based channel model accurately captures the statistical characteristics of the physical FSO channel in all three turbulence conditions. In each case, the KL divergence remains below 0.01 nats, with the lowest value reaching

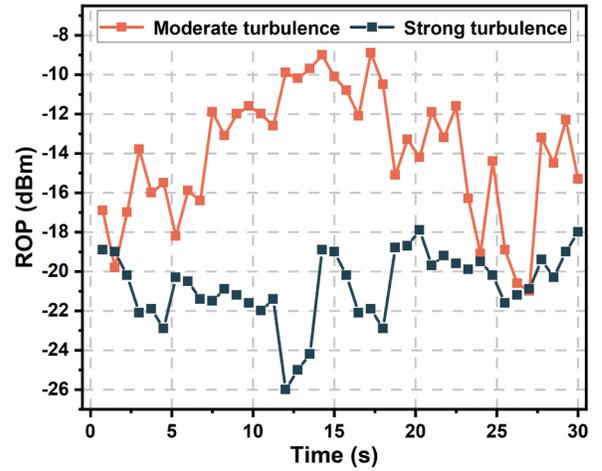

**Fig. 4.** Measured ROP of the FSO system.

0.0019 nats under weak turbulence.

In addition to amplitude distribution comparisons, we further evaluate the temporal behavior of the BiLSTM-based channel model by analyzing its output waveforms. Fig. 3(d)-(f) show time-domain waveform comparisons between the modeled and measured signals under the same three turbulence conditions. The generated waveforms show high similarity to the experimental measurements, further validating the model's ability to capture the temporal characteristics of the physical FSO channel.

*B. Result Analysis of TRFSO*

We conducted experimental validation under moderate and strong turbulence conditions by generating random phase screens to emulate atmospheric turbulence. For moderate turbulence, the parameters were set as $C_n^2 = 10^{-15} m^{-2/3}$, Fried coherence length $r_0 \approx 0.119 m$, transmission distance of 5 km, and an EDFA output power of 20 dBm. For strong turbulence, the parameters were $C_n^2 = 10^{-13} m^{-2/3}$, atmospheric coherence length $r_0 \approx 0.0075 m$, transmission distance of 5 km, and EDFA output power of 24 dBm. The received optical powers (ROP) under both turbulence levels are shown in Fig. 4. Due to atmospheric turbulence, the ROP exhibited significant fluctuations, with power decreasing as the turbulence strength increases.

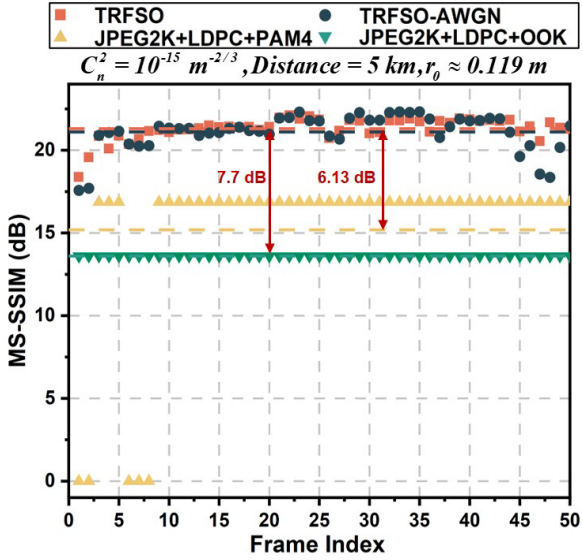

**Fig. 5.** MS-SSIM performance of different schemes at different turbulence realizations under moderate turbulence.

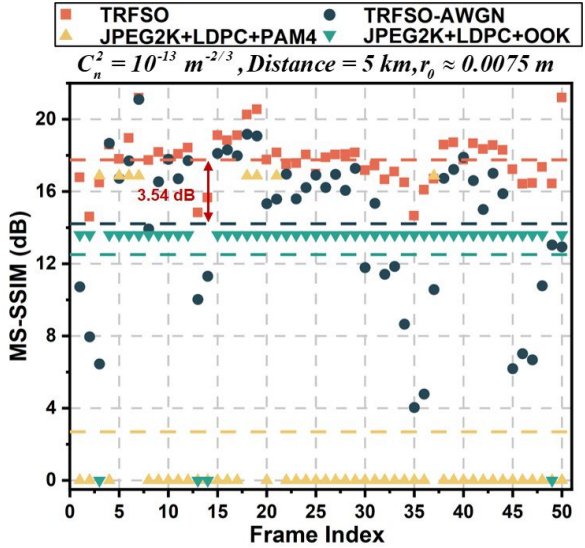

**Fig. 6.** MS-SSIM performance of different schemes at different turbulence realizations under strong turbulence.

In FSO communication, the substantial power losses and time-varying channel conditions pose significant challenges to reliable data transmission. These impairments highlight the need for highly robust systems that can maintain stable performance across varying turbulence levels. However, conventional separation-based FSO systems are susceptible to the cliff effect, leading to communication interruptions. To evaluate the performance of different transmission schemes, we conducted 50 random realizations for both moderate and strong turbulence.

As shown in Fig. 5, under moderate turbulence, the TRFSO system demonstrated a marked improvement in image transmission quality, achieving average MS-SSIM gains of 7.70 dB and 6.13 dB compared to the traditional FSO schemes based on PAM4 and OOK, respectively. The average values for each scheme are marked with dashed lines in the figure. This illustrates the TRFSO system's ability to deliver high-quality image transmission under favorable channel conditions.

Under strong turbulence, the traditional FSO communication system based on PAM4 experienced numerous transmission failures, successfully transmitting images in only 8 out of 50 realizations, as illustrated in Fig. 6. The TRFSO-AWGN scheme also exhibited substantial performance variations. In contrast, the TRFSO system maintained stable and high-quality image transmission. Compared to TRFSO-AWGN, TRFSO achieved an average MS-SSIM improvement of 3.54 dB, demonstrating its robustness in challenging atmospheric conditions. These results confirm that the TRFSO system can provide reliable and high-quality transmission across a wide range of turbulence levels, making it a promising solution for FSO communication.

We present visual comparisons of reconstructed images across different schemes, as shown in Fig. 7 and Fig. 8. Under moderate turbulence, the TRFSO system consistently yields high-quality reconstructed images with clear structural details and fewer distortions. The MS-SSIM values remain above 21 dB for most frames, representing a significant improvement over the PAM4-based and OOK-based traditional FSO systems. In the case of strong turbulence, the performance gap becomes more pronounced. The traditional PAM4-based system fails to recover the image in many realizations, while TRFSO-AWGN suffers from severe quality degradation. By contrast, our TRFSO system maintains stable visual quality, with most frames achieving MS-SSIM values above 16 dB and retaining key image features.

To evaluate the feasibility of the proposed TRFSO system in satellite communication (SatCom) scenarios, we analyze its performance under varying zenith angles and ground station altitudes, as illustrated in Fig. 9(a) and Fig. 9(b), respectively. Atmospheric turbulence is modeled using a phase screen generated by layering multiple turbulence layers corresponding to different turbulence strengths. These strengths are determined based on combined factors such as zenith angle, ground station altitude, and atmospheric structure parameters. This layered modeling approach allows for a more accurate representation of the turbulence characteristics encountered in realistic SatCom environments. The zenith angle refers to the angle between the line-of-sight to the satellite and the local vertical direction. As the zenith angle increases, the effective atmospheric propagation path lengthens, leading to an increase in both the transmission distance and the turbulence depth. Higher-altitude stations generally experience shorter turbulence paths and lower refractive index fluctuations, which leads to improved transmission quality. By incorporating these factors, our analysis provides a realistic assessment of the TRFSO

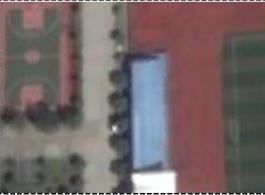

Fig. 7. Examples of visual comparison under moderate turbulence at $C_n^2 = 10^{-15} m^{-2/3}$ and fried coherence length $r_0 \approx 0.119m$.

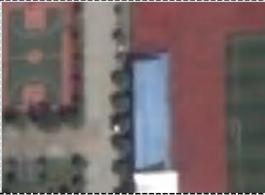

Fig. 8. Examples of visual comparison under strong turbulence at $C_n^2 = 10^{-13} m^{-2/3}$ and fried coherence length $r_0 \approx 0.0075m$.

system's performance under satellite-ground link conditions.

As shown in Fig. 9(a), the traditional FSO system exhibits the cliff effect as the zenith angle increases, where reliablecommunication cannot be sustained unless the modulation order is reduced, which comes at the expense of spectral efficiency and image quality. In contrast, transmission schemes based on JSCCM exhibit a more gradual degradation with increasing zenith angle and consistently outperform the traditional approach. It is worth noting that, at larger zenith angles, the TRFSO system achieves notably better performance compared to TRFSO-AWGN. Even at a zenith angle of 60°, TRFSO attains an MS-SSIM of 14.88 dB. These results confirm its potential to support robust image transmission in high-mobility and wide-angle SatCom scenarios.

In Fig. 9(b), a similar trend is observed with respect to ground station altitude. Traditional FSO systems again suffer from the cliff effect as altitude decreases and atmospheric turbulence intensifies. By contrast, the performance of TRFSO improves steadily with increasing altitude. At high altitudes, TRFSO achieves comparable performance to TRFSO-AWGN. At low altitudes, where turbulence is strongest and more variable, TRFSO significantly outperforms all baselines. These results demonstrate the robustness of the proposed model under complex channel conditions and highlight its suitability for deployment across a wide range of environments, including low-altitude and coastal regions where reliable SatCom remains challenging.

We conducted a comparison under the AWGN channel, as illustrated in Fig. 10. The encoder and decoder network structures and training hyperparameters in TRFSO-AWGN are kept the same as those used in TRFSO. This ensures that any performance differences can be attributed solely to the channel model, eliminating confounding factors related to network structure or training configuration. The TRFSO-AWGN system is trained on the AWGN channel. The TRFSO is trained under the proposed BiLSTM-based channel model. As expected, TRFSO-AWGN demonstrates better performance under the AWGN channel, outperforming TRFSO across the entire SNR range. By contrast, as evidenced by the results above, TRFSO achieves clear performance gains over TRFSO-AWGN in the FSO channel, owing to its robustness acquired through joint training with the turbulence-aware channel model. These results indicate that TRFSO-AWGN is more suitable for turbulence-free AWGN-like conditions, such as inter-satellite links, while TRFSO provides significant advantages in turbulence-impaired scenarios, such as space-to-ground links. This finding validates the critical role of accurate channel modeling in the design of JSCCM-based FSO systems. Moreover, the proposed TRFSO system introduces only a small increase in training complexity because of the channel model, while requiring no changes to the deployed encoder–decoder architecture and adding no extra computational cost during deployment. This makes it a scalable and cost-effective solution for real-world FSO communication systems.

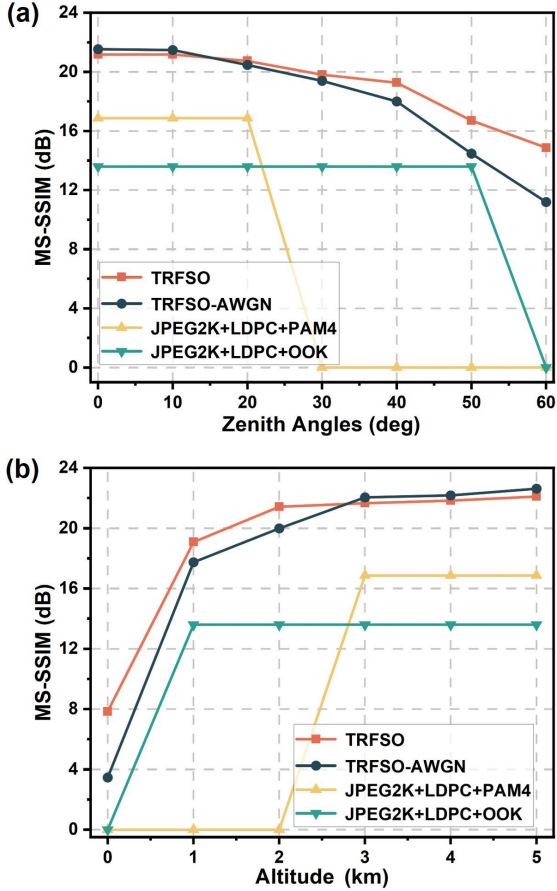

**Fig. 9.** MS-SSIM performance of different FSO schemes under varying (a) zenith angles and (b) ground station altitudes.

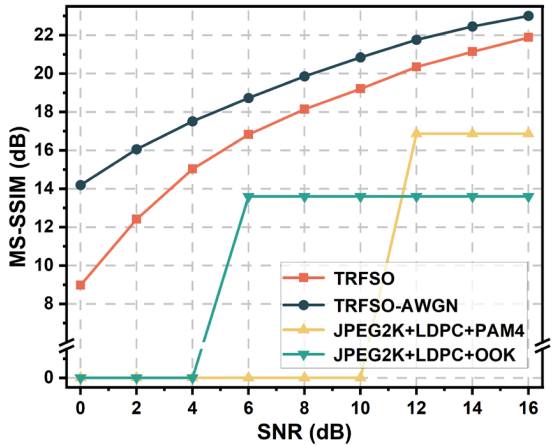

**Fig. 10.** Performance comparison under AWGN channel across varying SNR.

## V. CONCLUSION

In this work, we proposed TRFSO, an end-to-end FSO communication system designed for enhanced robustness under atmospheric turbulence. By integrating a BiLSTM-based channel model into the JSCCM framework, the system enables end-to-end optimization over non-Gaussian, time-varying FSO channels. The proposed channel model accurately emulates the impairments of FSO links, achieving a minimum KL divergence of 0.0019 in amplitude distribution matching. Extensive experimental results under varying turbulence intensities, zenith angles, and ground station altitudes demonstrate the superior adaptability of TRFSO. Under moderate turbulence, TRFSO sustains high image reconstruction quality, with average MS-SSIM levels exceeding 21.3 dB. Under strong turbulence, TRFSO achieves an average MS-SSIM improvement of 3.54 dB over the JSCCM model trained under the AWGN channel. Besides, TRFSO maintains consistently reliable performance across wide zenith angles and low altitude scenarios. These results highlight the importance of realistic channel modeling and validate TRFSO as a practical and scalable solution for reliable optical communication in terrestrial and SatCom scenarios.